# CausalCellSegmenter: Causal Inference inspired Diversified Aggregation Convolution for Pathology Image Segmentation

Dawei Fan[1] , Yifang Gao[1], Jiaming Yu[1], Yanping Chen[2,3], Wencheng Li[1], Chuancong Lin[1], Kaibin Li[1], Changcai Yang[1], Riqing Chen[1], Lifang Wei[1]

[1] College of Computer and Information Science, Fujian Agriculture and Forestry University, Fuzhou, China.
**dawei.fan@fafu.edu.cn**
[2] Department of Pathology, Clinical Oncology School of Fujian Medical University, Fuzhou, China.
[3] Fujian Cancer Hospital, Fuzhou, China.

**Abstract.** Deep learning models have shown promising performance for cell nucleus segmentation in the field of pathology image analysis. However, training a robust model from multiple domains remains a great challenge for cell nucleus segmentation. Additionally, the shortcomings of background noise, highly overlapping between cell nucleus, and blurred edges often lead to poor performance. To address these challenges, we propose a novel framework termed CausalCellSegmenter, which combines Causal Inference Module (CIM) with Diversified Aggregation Convolution (DAC) techniques. The DAC module is designed which incorporates diverse downsampling features through a simple, parameter-free attention module (SimAM), aiming to overcome the problems of false-positive identification and edge blurring. Furthermore, we introduce CIM to leverage sample weighting by directly removing the spurious correlations between features for every input sample and concentrating more on the correlation between features and labels. Extensive experiments on the MoNuSeg-2018 dataset achieves promising results, outperforming other state-of-the-art methods, where the mIoU and DSC scores growing by 3.6% and 2.65%.

## 1    Introduction

Over the past decade, deep learning has reached promising results for cell nucleus segmentation in the field of pathology image analysis [1–5]. For instances, the U-Net proposed by Ronneberger [1] and its improved version, U-Net++ by Zhou et al. [2],

---

Y. Gao and J. Yu contributed equally to this work.

with their unique encoder-decoder structure and skip connections [6], effectively integrate low-level to high-level semantic information of images, which greatly promotes the development of cell nucleus segmentation technology. Moreover, the introduction of the Transformer [7] architecture, with its unique attention mechanism [3], has optimized the processing of complex medical images, captured the intricate spatial relationships and feature hierarchies within images. Additionally, TransUNet [4] combines the image encoding capabilities of Convolutional Neural Network (CNN) with the deep contextual understanding of visual Transformers [8], bringing new perspectives and powerful potential to cell nucleus segmentation technology. However, the existing studies have shown limited performance due to issues such as background noise, highly overlapping between cell nucleus, and blurred edges, as demonstrated in Fig. 1. These shortcomings affect the accuracy and clarity of segmentation. In particular, when dealing with highly overlapping or closely spaced cell nucleus, the model performs poorly in distinguishing between cell nucleus and their surrounding tissue. Additionally, applying deep learning model to clinical practice hardly achieve promising segmentation performance due to the significant variations in lighting, resolution, and staining techniques of pathology images obtained from different hospitals or devices.

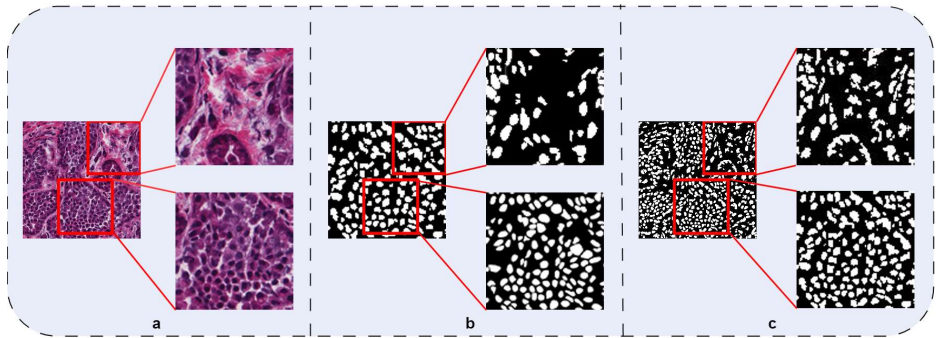

**Fig. 1.** Cell nucleus segmentation problem illustration. (a) denotes the micrographs image (b) denotes the ground truth and (c) denotes the predicted results obtained from Attention U-Net.

To tackle these challenges, we propose a novel framework termed CausalCellSegmenter, which combines Causal Inference Module (CIM) with Diversified Aggregation Convolution (DAC) techniques. The DAC module incorporates diverse downsampling features through a simple, parameter-free attention module (SimAM) [9], aiming to overcome the problems of false-positive identification and edge blurring, which significantly improves the accuracy and clarity of cell nucleus segmentation. Furthermore, we design CIM to mitigate the decline of segmentation performance due to the data distribution variations from different hospitals and devices inspired by Zhang et al [10]. In particular, the CIM leverages sample weighting to directly remove the spurious correlations between features for every input sample and concentrate more on the correlation between features and labels. In summary, the key contributions of our work are as follows: (1) The DAC module is designed by optimizing various types of semantic information [11] for feature fusion, while reducing the occurrence of false-positive predictions. (2) We introduce CIM by adopting a causal inference learning strategy and dynamically adjusting the sample weights, which enhances the recognition ability of cell nucleus features and effectively reduces the negative impact of spurious correlations. The comprehensive experiments show that our CausalCellSegmenter provides promising results, that is superior to other state-of-the-art methods. These advances have provided a new direction for addressing the domain shift [12] problem in pathology image analysis, offered dependable methods to improve the accuracy of pathology image segmentation.

## 2 Method

In this section, we first introduce an overview of our CausalCellSegmenter. Then, we detail each part of the CausalCellSegmenter.

### 2.1 Overview

The overview of our CausalCellSegmenter is given in Fig. 2. The framework begins with a micrographs image input, which undergoes five layers of downsampling by CNN and Mobile Inverted Bottleneck Convolution (MBConv) [13], producing multi-level feature maps from low to high levels. After each downsampling stage, the resulting feature maps are combined using DAC in the next downsampling process. Simultaneously, the attention mechanism of the Transformer will enrich the feature maps generated by downsampling the first layer of the CNN, resulting in a more integrated set of feature maps. For the final generated feature maps, CIM is used to learn the sample weights for every sample and compute the weighted loss. Afterwards, the feature maps are fused progressively using skip connections and a convolutional decoder, similar to the U-Net structure. Finally, we leverage the softmax function to obtain the cell nucleus segmentation map.

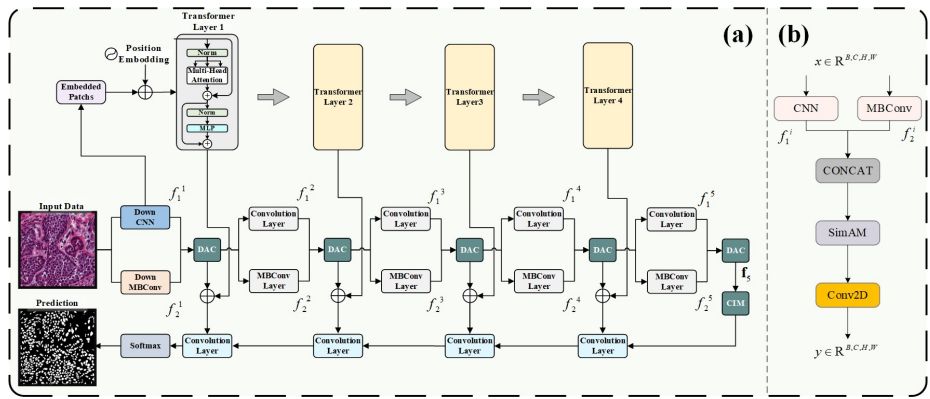

**Fig. 2.** The overview of CausalCellSegmenter (a) with DAC module (b).

## 2.2 Diversified Aggregation Convolution (DAC)

To enhance the precision of cell nucleus segmentation, particularly in addressing segmentation challenges arising from issues such as blurred edges and background noise, we introduce an effective Diversified Aggregation Convolution (DAC) module. The DAC is specifically designed to integrate multi-scale feature maps [14] with rich semantic information. It optimizes the feature extraction process using CNN and MBConv and focuses on regions of interest by a simple, parameter-free attention module (SimAM) [9], achieving precise discrimination between nucleus and non-nucleus.

As demonstrated in Fig. 2(b), the input data undergoes a downsampling $f_1^i$ and $f_2^i$ after using CNN and MBConv, where $f_1^i$ represents the i-th layer downsampling using CNN and $f_2^i$ represent the corresponding layer using MBConv. Afterwards, we concatenate $f_1^i$ and $f_2^i$ by learned weights $k_1^i$ and $k_2^i$. The obtained features are then fed into SimAM, which aims to accurately evaluate and determine the most important features of semantic information in the cell nucleus segmentation. Subsequently, these features from SimAM are downsampled using two-dimensional convolution (Conv2D). Hence, the DAC can be represented as follows:

$$f_i = Conv2D(SimAM([k_1^i f_1^i, k_2^i f_2^i]))$$ (1)

Where [·] denotes the CONCAT operation and $f_i$ denotes i-th layer feature maps after DAC. We initialize $k_1^i$ and $k_2^i$ as 1.0.

## 2.3 Causal Inference Module (CIM)

To address data heterogeneity problem, we follow the StableNet [10] framework and design a causal inference module (CIM). We obtain the feature maps $f_5$ from the 5-th layer DAC as shown in Fig. 2. To remove spurious correlations between features, we leverage Random Fourier Features (RFF) extractor and sample weighting for segmentation task followed by our previous work [15]. Specifically, we use $A$ and $B$ represent the feature variable in the feature map from $f_5$. As Frobenius norm of the partial cross-covariance matrix $\|\sum_{AB}\|_F^2$ tends to zero, the two variables $A$ and $B$ are independent [10]. Hence, the partial cross-covariance matrix be:

$$\sum_{AB} = \frac{1}{n-1}\sum_{i=1}^{n}\left[\left(u\left(A_i\right)-\frac{1}{n}\sum_{j=1}^{n}u\left(A_j\right)\right)^T \cdot \left(v\left(B_i\right)-\frac{1}{n}\sum_{j=1}^{n}v\left(B_j\right)\right)\right]$$ (2)

Where $u(\cdot)$ and $v(\cdot)$ represent the RFF mapping functions, $n$ represents the number of input images.

We use $\boldsymbol{w} \in \mathbb{R}_+^n$ to represent the sample weights and $\sum_{i=1}^{n}w_i = n$ in the Weight Learner as shown in Fig. 3. After weighting, the partial cross-covariance matrix for variables $A$ and $B$ is as follows:

$$\sum_{AB};w = \frac{1}{n-1}\sum_{i=1}^{n}\left[\left(w_i u\left(A_i\right)-\frac{1}{n}\sum_{j=1}^{n}w_j u\left(A_j\right)\right)^T \cdot \left(w_i v\left(B_i\right)-\frac{1}{n}\sum_{j=1}^{n}w_j v\left(B_j\right)\right)\right]$$ (3)

$A_i$ and $B_i$ ($i \in [1, n]$) are sampled from the feature distribution of $A$ and $B$.

The objective function of weight learner is:

$$\boldsymbol{w}^* = \underset{\boldsymbol{w} \in \Delta_n}{arg\,min} \sum_{1 \le i \le j \le m} \|\Sigma_{AB}; w\|_F^2 \tag{4}$$

Where $m$ represents the number of features and $\Delta_n = \{\boldsymbol{w} \in \mathbb{R}_+^n | \sum_{i=1}^n w_i = n\}$. Hence, weighting training samples with the optimal $\boldsymbol{w}^*$ can mitigate the dependence between features and consequently remove spurious correlations between features to the greatest extent.

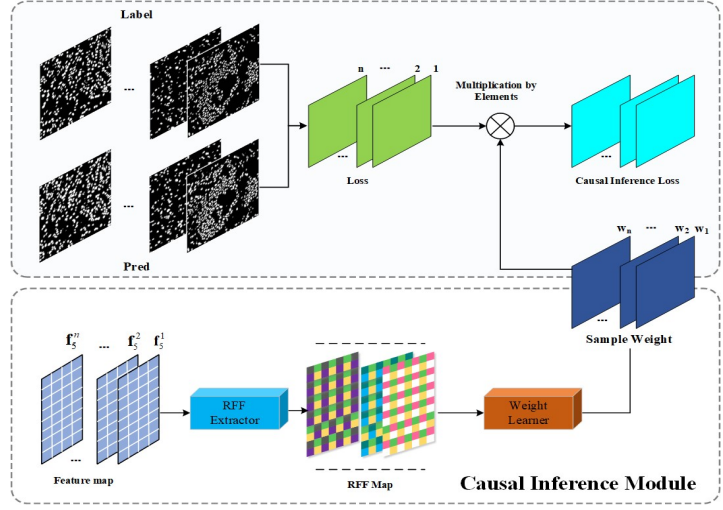

**Fig. 3.** Detailed learning procedure of Causal Inference Module.

Finally, the causal inference loss is as follows:

$$L_{CIM} = \sum_{i=1}^n w_i \otimes f(L_{CE}(X_i, y_i)) \tag{5}$$

Where $i$ represents the size of batch size. $L_{CE}(\cdot, \cdot)$ is the cross-entropy loss and $\otimes$ represents element-wise multiplication. $f(\cdot)$ represents the flattening operation.

Furthermore, we leverage dice loss and focal loss [16] to optimize our framework simultaneously. The dice loss is a common loss in the segmentation tasks. And the focal loss aims to solve the problem of category imbalance by reducing the impact of easily classifiable samples and enhancing the challenging ones:

$$L_{FL} = -\alpha_t(1 - p_t)^\gamma log(p_t) \tag{6}$$

Where $\alpha_t$ is a balancing coefficient for cell nucleus and background. $p_t$ represents the predicted probability for the correct class. $\gamma$ is a tuning factor to adjust the weights of background. In this paper, we set $\alpha_t$ as 0.8 and $\gamma$ as 2.0.

Hence, the total loss is:

$$L = L_{\text{CIM}} + \lambda L_{\text{Dice}} + (1 - \lambda)L_{FL} \tag{7}$$

Where $\lambda$ represents the coefficient of dice loss and $(1 - \lambda)$ represents the coefficient of focal loss. We set $\lambda$ as 0.5.

## 3    Experiments

### 3.1    Dataset and Experimental Details

In the task of cell nucleus segmentation, which requires precise pixel-level annotation, manual labeling is time-consuming and labor-intensive. Therefore, to evaluate various models effectively, we chose a smaller dataset, MoNuSeg-2018 [17]. This dataset comprises tissue sample images from tumor patients across different organs, meticulously annotated across multiple medical institutions. These images are 40x magnified H&E stained tissue sections originating from the Cancer Genome Atlas (TCGA) database. The dataset exhibits significant domain variations due to the diversity of organs, patients, and staining protocols across hospitals. For consistency and reproducibility, all experiments were conducted using the PyTorch 1.13.0 framework on an NVIDIA RTX2080Ti GPU and an i7-13700k CPU with 64GB RAM. The model is trained using the AdamW optimizer with a cosine annealing strategy [18] to adjust the learning rate, which is set to 1e-3. In addition, the model training batch size is set to 8, and the entire training period is set to 400 epochs. To prevent overfitting, an early stopping strategy is adopted. The experimental images are resized to 224x224 and data augmentation techniques are applied, including horizontal flipping, rotation, gaussian blur, color intensity enhancement, and random cropping.

### 3.2    Evaluation metrics

In this work, we use two core metrics to evaluate model performance: the Dice Coefficient (DSC) and the Mean Intersection over Union (mIoU) [17]. DSC is suitable for binary image segmentation and measures sample similarity, while mIoU assesses the overlap between predicted and actual areas. These metrics together provide a precise evaluation of performance, verifying the model's effectiveness in complex image segmentation tasks.

### 3.3    Experimental Results

Comparative experiments are conducted with five state-of-the-art (SOTA) models. The results, presented in Table 1, demonstrate that our framework outperforms Attention U-Net in mIoU and DSC by 3.6% and 2.65%, respectively. In particular, when compared to Transformer-based U-Net models such as TransUNet and Swin-UNet, our model shows even more improvements, with increases of 8.87% and 10.26% in mIoU, and 6.6% and 7.75% in DSC, respectively. These medical image segmentation evaluation metrics indicate that CausalCellSegmenter can achieve the highest scores.

Cell nucleus segmentation is essential in pathology image analysis. To demonstrate the superiority of our CausalCellSegmenter, we illustrate the segmentation comparison results with other advanced models. According to the visualization results presented in Fig. 4, CausalCellSegmenter outperforms other models in generating precise cell nucleus segmentation results and reducing false positives. This demonstrates the efficiency and accuracy of CausalCellSegmenter for cell nucleus segmentation from pathology images, particularly in distinguishing closely adjacent nucleus and accurately identifying cell nucleus boundaries.

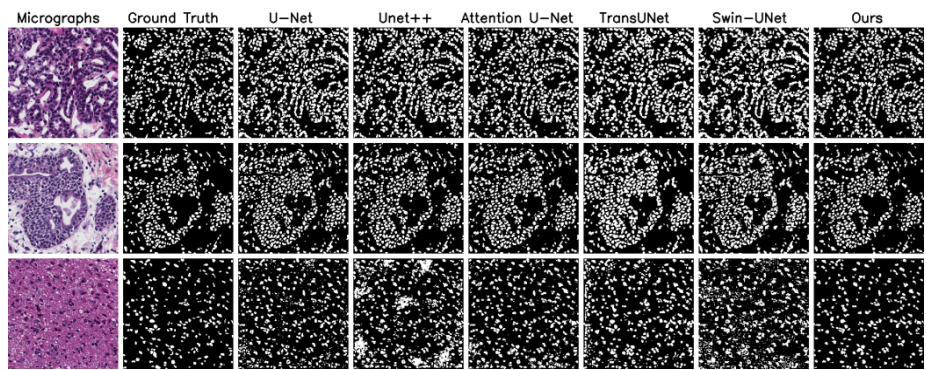

**Fig. 4.** Comparison of our framework and other SOTA methods' results visualization on the MoNuSeg-2018.

**Table 1.** Performance comparison: CausalCellSegmenter versus SOTA on MoNuSeg-2018.

| Method | mIoU(%) | DSC(%) |
|---|---|---|
| Swin-UNet [5] | 58.74±5.65 | 73.85±4.67 |
| TransUNet [4] | 60.13±4.84 | 75.00±3.82 |
| U-Net [1] | 62.82±5.00 | 77.06±3.78 |
| U-Net++ [2] | 63.89±9.00 | 77.59±7.50 |
| Attention U-Net [2] | 65.40±5.70 | 78.95±4.23 |
| Ours | **69.00±3.80** | **81.60±2.66** |

### 3.4 Ablation Study

We further conduct ablation studies to evaluate the effectiveness of crucial components in our CausalCellSegmenter. As documented in Table 2, Backbone represents the combination of CNN downsampling and Transformer, which shows limited performance. For training model with Backbone+DAC, as shown in Table 2, Backbone+DAC outperforms Backbone with a growth of 1.83% and 1.33% in mIoU and DSC, respectively. The promising results demonstrate that the DAC improves the accuracy of cell nucleus edges recognition by optimizing various types of semantic information [11] for feature fusion, while reducing the occurrence of false-positive predictions. Meanwhile, for training model with Backbone+CIM, it outperforms Backbone, increasing mIoU and DSC by 1.55% and 1.13%, respectively. This demonstrate that CIM can effectively remove spurious correlations between features and concentrate more on features that contribute to cell nucleus segmentation results. Besides, when performing Backbone+DAC+CIM on our model, there is an improvement compared to the Backbone, where the mIoU and DSC gain a growth of 2.25% and 1.63%, respectively. These results demonstrate that our CausalCellSegmenter can enhance the performance of cell nucleus segmentation task in the field of pathology image analysis.

**Table 2.** Quantitative evaluation of the proposed crucial modules in CausalCellSegmenter.

| CausalCellSegmenter | | | | |
| Backbone | DAC | CIM | mIoU(%) | DSC(%) |
| --- | --- | --- | --- | --- |
| ✓ | | | 66.75±4.66 | 79.97±3.32 |
| ✓ | | ✓ | 68.30±4.11 | 81.10±2.89 |
| ✓ | ✓ | | 68.58±3.92 | 81.30±2.73 |
| ✓ | ✓ | ✓ | 69.00±3.80 | 81.60±2.66 |

## 4 Conclusion

In this paper, focusing on the challenges of domain shifts and blurred edges in the cell nucleus segmentation tasks, we propose a novel CausalCellSegmenter that combines CIM with DAC. Compared to other methods, our CausalCellSegmenter can effectively improve the results of cell nucleus segmentation by removing spurious correlations between features, and focusing more on the correlations between features and labels. Meanwhile, it improves the accuracy of cell nucleus edges recognition by optimizing various types of semantic information for feature fusion, reducing the occurrence of false-positive predictions. The extensive experiments demonstrate that our CausalCellSegmenter can effectively alleviate the issues in the cell nucleus segmentation and achieves a robust performance.